\documentclass{article}
\usepackage[margin=2.5cm]{geometry}
\usepackage{graphicx} 
\usepackage{amsmath}
\usepackage[capitalise]{cleveref}
\usepackage{biblatex}
\usepackage{multirow} 
\usepackage{booktabs} 
\usepackage{caption} 
\usepackage{wasysym} 

\usepackage{qcircuit}

\newcommand{\controlu}{*-=[][F]{\phantom{\bullet}}}
\newcommand{\ctrlu}[1]{\controlu \qwx[#1] \qw}

\newcommand{\multistate}[2]{*+{\hphantom{#2}} \POS[0,0].[#1,0] !C *{#2} \POS[0,0].[#1,0] \drop\frm{}}
\newcommand{\ghoststate}[1]{*+{\hphantom{#1}} }
\newcommand{\ccteq}[1]{\multistate{#1}{=}}
\newcommand{\ccteqg}{\ghoststate{=}}
\newcommand{\csl}{{\ \; \backslash }}

\usepackage{tabularx} 
\usepackage[table]{xcolor} 
\usepackage{braket} 
\usepackage[table]{xcolor} 
\definecolor{gr}{HTML}{C0C0C0}  
\usepackage{hhline}

\title{Beyond Quantum Shannon: Circuit Construction for General n-Qubit Gates Based on Block ZXZ-Decomposition}

\author{Anna M. Krol, Zaid Al-Ars \\ Quantum \& Computer Engineering Dept. \\Delft University of Technology, Delft, The Netherlands}

\begin{document}

\maketitle

\begin{abstract}
This paper proposes a new optimized quantum block-ZXZ decomposition method~\cite{art:blockzxz,art:unifiedapproach,art:biunimodular} that results in more optimal quantum circuits than the quantum Shannon decomposition (QSD)~\cite{art:synthesisqlogic}, which was introduced in 2006 by Shende et al. 
The decomposition is applied recursively to generic quantum gates, and can take advantage of existing and future small-circuit optimizations. Because our method uses only one-qubit gates and uniformly controlled rotation-Z gates, it can easily be adapted to use other types of multi-qubit gates.
With the proposed decomposition, a general 3-qubit gate can be decomposed using 19 CNOT gates (rather than 20). For general $n$-qubit gates, the proposed decomposition generates circuits that have $\frac{22}{48}4^n - \frac{3}{2}2^n +\frac{5}{3}$ CNOT gates, which is less that the best known exact decomposition algorithm by $(4^{n-2} -1)/3$ CNOT gates.
\end{abstract}

\section{Introduction}
To execute a quantum algorithm, a series of unitary operations (gates) and non-unitary operations (measurements) are applied to quantum bits (qubits) in a quantum circuit. The complexity of a quantum algorithm can be described as the number of gates, the number of qubits or the length of the critical path (depth) of the circuit. 

Physically, a qubit is a quantum-mechanical system that can store quantum information, such as superconducting qubits~\cite{art:superconductingqubits}, trapped ions~\cite{art:trappedions} or spin qubits~\cite{art:spinqubits}. Applying a quantum gate means manipulating the state of the qubit in a controlled way. Exactly which gate operations are possible depends on the qubit technology and the implementation~\cite{art:experimentalcomparison}. 

To run arbitrary quantum operations on real quantum hardware, the unitary operator (matrix) needs to be translated into elementary (native) gate operations. This is by no means a trivial task, and the focus of much research over the years into methods for performing such translation using quantum gate decomposition.

An important target of gate decomposition methods is to minimize the number of two-qubit gates required to implement a given unitary matrix. This is essential, because the two-qubit gates require qubit connectivity and mapping, and the execution time and error-rates of two-qubit gates are an order of magnitude worse than for single qubit gates in current quantum hardware~\cite{art:experimentalcomparison}.  
 
It has been proven that any exact decomposition of an arbitrary $n$-qubit gate requires at least $\frac{1}{4}(4^n-3n-1)$ CNOT gates~\cite{art:minimalunivtwoq}.

Approximate decomposition algorithms such as~\cite{art:approachingthelimit, art:numericalanalysis,art:qfast} can be used to decompose arbitrary quantum gates with (almost) the minimum number of CNOT gates and little accuracy loss, at the cost of excessive runtime of the search algorithm: decomposition of a five qubit gate can take at least several hours. These methods are therefore not suitable for bigger gates or for applications where classical compile time is relevant for the performance of the algorithm~\cite{art:efficientparams}.

In contrast, exact decomposition methods are much faster, and for one- and two-qubit gates also achieve the minimum CNOT count.
One-qubit gates do not require any CNOTs and can be decomposed into a sequence of three rotation gates~\cite{art:elementarygates}. Arbitrary two-qubit gates can be decomposed into three CNOTs using the methods described in
~\cite{art:optimalquantumcircuitsforgen2q,art:minimalunivtwoq,art:universalqcirfor2q,art:universalquantumcirc}, which also show that less CNOTs are necessary when the gate meets certain conditions. 
For arbitrary three-qubit gates, there is no algorithm that results in the minimum 14 CNOTs, but algorithms do exist that can decompose them into 64~\cite{art:efficientdecompquantumgates}, 40~\cite{art:3Qin40}, 26~\cite{art:3Qin26} or 20~\cite{art:synthesisqlogic} CNOTs. 

For quantum gates of arbitrary size, the decomposition methods have drastically improved since 1995, when Barenco et al.~\cite{art:elementarygates} showed that any unitary operator on $n$ qubits can be constructed using at most $O(n^{3}4^n)$ two-qubit gates. This decomposition method used the standard QR decomposition based on Givens rotations~\cite{art:reducingqops}, and the CNOT count has been improved over the years by use of Gray codes and gate cancellations to $O(1/2\cdot 4^n)$ CNOT gates~\cite{art:palindrometrans, art:synthesisqlogic,art:efficientdecompquantumgates}. 
Another approach to unitary decomposition has been to use Cosine Sine Decomposition (CSD)~\cite{art:histandgenCSD,book:matrixcomp,art:rudimentarycomp,art:quantumcircgenmultigates}. This was combined with Singular Value Decomposition (SVD) in 2006~\cite{art:synthesisqlogic} to construct Quantum Shannon Decomposition (QSD). With QSD, an $n$-qubit unitary gate can be decomposed into at most $(23/48)\cdot 4^n - (3/2) \cdot 2^n + (4/3)$ CNOTs. 
More recently, the Khaneja-Glaser decomposition~\cite{art:khanejaglaser} was used in~\cite{art:notnearoptimalkgg} to construct a decomposition method that can decompose unitary operations using $(21/16)\cdot 4^n - 3(n\cdot 2^{n-2}+2^n)$ CNOT gates.

In this paper, we show the design and construction of a new unitary decomposition method based on block-ZXZ decomposition~\cite{art:blockzxz,art:unifiedapproach,art:biunimodular}, that uses demultiplexing and optimizations similar to quantum Shannon decomposition~\cite{art:synthesisqlogic}. 
The contributions of this paper are as follows.
\begin{itemize}
    \item We show how to decompose an arbitrary $n$-qubit gate into at most $(22/48)\cdot 4^n - (3/2)\cdot 2^n + (5/3)$ CNOT gates. This is $(4^{n-2} -1)/3$ less than the best previously published work~\cite{art:synthesisqlogic}.
    \item More specifically, we can construct a general three-qubit operator with at most 19 qubits, which is currently the least known for any exact decomposition method.
\end{itemize}

An overview of the CNOT count for the proposed method compared to previously published unitary decomposition algorithms is given in~\cref{tab:comparedecomps}.

\begin{table}[h]
\caption{Number of CNOT gates resulting from unitary decomposition by the proposed decomposition compared to previously published algorithms and the theoretical lower bound. The results of this paper are shown in bold.} \label{tab:comparedecomps}
\begin{tabular}{lllllllll} \toprule
Number of qubits                & 1          & 2          & 3           & 4           & 5            & 6             & $n$                                                   \\ \midrule
Original QR decomp.~\cite{art:elementarygates,art:reducingqops}                &            &            &             &             &              &               & $O(n^3\cdot 4^n)$                                     \\
Improved QR decomp.~\cite{art:knill1995}                &            &            &             &             &              &               & $O(n\cdot 4^n)$                                        \\
Palindrome transform~\cite{art:palindrometrans,art:synthesisqlogic}          &            &            &             &             &              &               & $O(n\cdot 4^n)$                                        \\ \midrule
Givens rotations (QR)~\cite{art:efficientdecompquantumgates}                & 0          & 4          & 64          & 536         & 4156         & 22618         & $\approx 8.7 \cdot 4^n $                             \\
Iterative disentangling (QR)~\cite{art:synthesisqlogic}           & 0          & 8          & 62          & 344         & 1642         & 7244          & $2\cdot 4^n - (2n+3)\cdot 2^n + 2^n $                 \\
KG Cartan decomp.~\cite{art:notnearoptimalkgg}         & 0          & 3          & 42          & 240         & 1128         & 4896          & $(21/16)\cdot 4^n - 3(n\cdot 2^{n-2}+2^n)$            \\
Original CSD~\cite{art:rudimentarycomp,art:effunidecomp}                            & 0          & 14         & 92          & 504         & 2544         & 12256         & $(1/2)\cdot n \cdot 4^n - (1/ 2) \cdot 2^n$           \\
CSD~\cite{art:quantumcircgenmultigates}                   & 0          & 8          & 48          & 224         & 960          & 3968          & $4^n-2\cdot2^n$                                       \\
CSD (optimized)~\cite{art:decompofgeneralquantumgates}              & 0          & 4          & 26          & 118         & 494          & 2014          & $(1/2) \cdot 4^n-(1/2)\cdot 2^n - 2$                  \\
QSD (base)~\cite{art:synthesisqlogic}                      & 0          & 6          & 36          & 168         & 720          & 2976          & $(3/4)\cdot 4^n - (3/2) \cdot 2^n $                   \\
\textbf{Block-ZXZ}~\cite{art:blockzxz}                & 0          & 6          & 36          & 168         & 720          & 2976          & $(3/4)\cdot 4^n - (3/2) \cdot 2^n $                   \\
QSD (optimized)~\cite{art:synthesisqlogic}                 & 0          & 3          & 20          & 100         & 444          & 1868          & $(23/48)\cdot 4^n - (3/2)\cdot 2^n + (4/3) $          \\
\textbf{Proposed decomposition} & \textbf{0} & \textbf{3} & \textbf{19} & \textbf{95} & \textbf{423} & \textbf{1783} & $\mathbf{(22/48)\cdot 4^n - (3/2)\cdot 2^n + (5/3) }$ \\ \midrule
Theoretical lower bounds        & 0          & 3          & 14          & 61          & 252          & 1020          & $(1/4) \cdot (4^n-3n-1)$      \\ \bottomrule          
\end{tabular}
\end{table}

The rest of the paper is organized as follows. We start with the notation and gate definitions in \cref{sec:notation}. Then in \cref{subsec:uniformlycontrolled}, we show the decomposition of uniformly controlled rotations. \cref{sec:decomposition} continues  with the full decomposition. The optimizations and the resulting gate count are shown in \cref{sec:optimization}. The paper ends with the conclusion in \cref{sec:conclusion}.

\section{\label{sec:notation} Notation and gate definitions}
This section introduces the mathematical notation and gate definitions used in this paper.

\subsection{Mathematical operations}
The conjugate transpose of a matrix is represented with $\dagger$ (i.e. the conjugate transpose of matrix $U$ is $U^\dagger$). 
Reversible quantum operations (gates) can be fully represented as unitary matrices, for which $U^\dagger = U^{-1}, U U^\dagger = I$, where $I$ is the identity matrix. 

The Kronecker product of two matrices is written as $\otimes$. The Kronecker product of ($n \times m$) matrix $A$ and ($p \times q$) matrix $B$ is the ($pm \times qn$) block matrix:
\[
A \otimes B = \begin{bmatrix} a_{11}B & \cdots & a_{1n}B \\ \vdots & \ddots & \vdots \\ a_{m1}B & \cdots & a_{mn}B \end{bmatrix}
\]
The Kronecker sum of two matrices is written as $\oplus$. The Kronecker sum of ($n \times m$) matrix $A$ and ($p \times q$) matrix $B$ is the ($(m+p) \times (n+q)$) block matrix:
\[
A \oplus B = \begin{bmatrix} A & 0 \\ 0 & B \end{bmatrix}
\] where the zeros are zero matrices.

\subsection{Elementary gates}
The elementary quantum operations used in this paper are listed below. These are part of the well-established and widely used set presented in~\cite{art:elementarygates}. The definition for the $R_z$ gate is the same as the one used in the quantum Shannon decomposition~\cite{art:synthesisqlogic}.
\begin{itemize}
    \item Identity gate: I $ = \begin{array}{lcr} \Qcircuit @C=1em @R=.7em {
& \gate{I} & \qw } \end{array} = \begin{bmatrix} 1 & 0 \\ 0 & 1 \end{bmatrix}$
    \item Rotation-X gate: \\$ R_x(\theta) = \begin{array}{lcr} \Qcircuit @C=1em @R=.7em {
    & \gate{R_x} & \qw } \end{array} = \begin{bmatrix} cos(\theta / 2) & i \cdot sin(\theta /2) \\ i \cdot sin(\theta /2) & cos(\theta /2) \end{bmatrix} $
    \item Rotation-Y gate: \\ $ R_y(\theta) = \begin{array}{lcr} \Qcircuit @C=1em @R=.7em {
    & \gate{R_y} & \qw } \end{array} = \begin{bmatrix} cos(\theta / 2) &  sin(\theta /2) \\ - sin(\theta /2) & cos(\theta /2) \end{bmatrix} $
    \item Rotation-Z gate: \\$ R_z(\theta) = \begin{array}{lcr} \Qcircuit @C=1em @R=.7em {
& \gate{R_z} & \qw } \end{array} = \begin{bmatrix} e^{-i\theta/2} & 0 \\ 0 & e^{i\theta/2} \end{bmatrix} $
    \item Hadamard gate: H $=  \begin{array}{lcr} \Qcircuit @C=1em @R=.7em {
& \gate{H} & \qw } \end{array} = \frac{1}{\sqrt{2}} \begin{bmatrix} 1 & 1 \\ 1 & -1 \end{bmatrix} $
    \item Controlled-not gate: \\CNOT $ = \begin{array}{lcr} \Qcircuit @C=1em @R=.7em {
& \ctrl{1} & \qw \\ & \targ & \qw } \end{array} = \begin{bmatrix}1 & 0 & 0 & 0 \\ 0 & 1 & 0 & 0 \\ 0 &  0 & 0 & 1 \\ 0 & 0 & 1 & 0 \end{bmatrix} $
    \item Z-gate: Z $ =  \begin{array}{lcr} \Qcircuit @C=1em @R=.7em {
& \gate{Z} & \qw } \end{array} = \begin{bmatrix} 1 & 0 \\ 0 & -1 \end{bmatrix} $
    \item Controlled-Z gate: \\CZ $ = \begin{array}{lcr} \Qcircuit @C=1em @R=.7em {
& \ctrl{1} & \qw \\ & \gate{Z} & \qw } \end{array} = $ $\begin{array}{lcr} \Qcircuit @C=1em @R=1em {
& \control{1} \qw & \qw \\ & \ctrl{-1} & \qw } \end{array} =  \begin{bmatrix}1 & 0 & 0 & 0 \\ 0 & 1 & 0 & 0 \\ 0 &  0 & 1 & 0 \\ 0 & 0 & 0 & -1 \end{bmatrix}$
\end{itemize}

\subsection{Generic gates}
The following generic gates are used in the decomposition and their circuit representation are listed below.
\begin{itemize}
    \item Generic single-qubit unitary gate: \\
$ U(2) = \begin{array}{lcr} \Qcircuit @C=1em @R=.7em { & \gate{U} & \qw } \end{array} $
\item Generic multi-qubit unitary gate: \\
$ U(n) = \begin{array}{lcr}\Qcircuit @C=1em @R=.7em {
\csl & \gate{U} & \qw }\end{array} $ where the backslash is used to show that the wire carries an arbitrary number of qubits.
\item Controlled arbitrary (multi-qubit) gates: \\
$ \begin{array}{lcr}  \Qcircuit @C=1em @R=.7em { & \ctrl{1} & \qw \\  \csl & \gate{U} & \qw }\end{array}= \begin{bmatrix} I & 0 \\ 0 & U \end{bmatrix} $, gate $U$ is only applied if the control qubit is in state $\ket{1}$.
\item Quantum multiplexor: 
$ \begin{array}{lcr}  \Qcircuit @C=1em @R=.7em { &  & \ctrlu{1} & \qw \\ & \csl & \gate{U} & \qw }\end{array} = U_1 \oplus U_2 = \begin{bmatrix} U_1 & 0 \\ 0 & U_2 \end{bmatrix} $, gate $U_1$ is applied if the control qubit is in state $\ket{0}$, gate $U_2$ is applied if the control qubit is in state $\ket{1}$.
\item Uniformly controlled rotation gate: \\
$ \begin{array}{lcr}  \Qcircuit @C=1em @R=.7em { & \csl & \ctrlu{1} & \qw \\ & & \gate{R_a} & \qw }\end{array}$, a different rotation around axis $a$ is applied depending on the state of the control qubits.
\end{itemize}

\section{Decomposing uniformly controlled rotations} \label{subsec:uniformlycontrolled}
This section shows the decomposition for one of the main building blocks resulting from our method; the uniformly controlled rotation gates. 
These gates will be decomposed using the method from~\cite{art:quantumcircgenmultigates}. 

The uniformly controlled rotation gates that are used in our decomposition method are always uniformly controlled $R_z$ gates applied to the first qubit. 
The matrix representation of such a gate follows from the general matrix representation of a uniformly controlled $R_z$ gate with $k$ controlling qubits, that is applied to the \textit{last} qubit ($k+1$). The matrix representation of such a gate is:
\begin{align}
F_{k+1}^k\left(R_z(\alpha_j)\right) = \begin{bmatrix} R_z(\alpha_1) & & \\ & \ddots & \\ & & R_z(\alpha_{2^k}) \end{bmatrix} \label{eq:Fka}  
\end{align}

The matrix corresponding to a uniformly controlled $R_z$ gate applied to the \textit{first} qubit is $F_1^k (R_z) = (D\oplus D^\dagger)$, where D is a ($2^k \times 2^k$) diagonal matrix~\cite{art:synthesisqlogic} consisting of the upper left entries of the matrices for $R_z(\alpha_j)$. 
With $R_z$ defined as in \cref{sec:notation}, the angles $\alpha_j$ can be calculated as:
\begin{align}
        e^{-i/2  \cdot \makebox{\normalsize\ensuremath{ \alpha_j}}} &= diag(D_j),~j = {1,\,\dotsb , 2^k}
\end{align}
where $diag(D_j)$ is the $j^\mathrm{th}$ diagonal element of $D$ and $\alpha_j$ is the $j^\mathrm{th}$ angle $\alpha$ in \cref{eq:Fka}.

This gate can be implemented by an alternating sequence consisting of $2^k$ CNOTs and $2^k$ single qubit rotation gates applied to the target qubit. The CNOT controls are determined using a sequence based on the binary reflected Gray code~\cite{pat:graycode}, calculated as follows: the position of the control for the $l^{\mathrm{th}}$ CNOT matches the position where the $l^{\mathrm{th}}$ and the $(l+1)^{\mathrm{th}}$ bit strings of the Gray code differ.

\begin{figure}[h]
\centering
\leavevmode
\Qcircuit @C=0.3em @R=.7em {
& \ctrlu{1}   &\qw    & \ccteq{3}
& &\qw 
& \qw & \qw & \qw & \qw & \qw & \qw & \qw & \ctrl{3}
& \qw & \qw & \qw & \qw & \qw & \qw & \qw & \ctrl{3}&\qw\\
&\ctrlu{1}    &\qw    & \ccteqg
& &\qw 
& \qw & \qw & \qw & \ctrl{2} & \qw & \qw & \qw & \qw
& \qw & \qw & \qw & \ctrl{2} & \qw & \qw & \qw & \qw &\qw\\
& \ctrlu{1}   &\qw    &\ccteqg
& &\qw 
& \qw & \ctrl{1} & \qw & \qw & \qw & \ctrl{1} & \qw & \qw
& \qw & \ctrl{1} & \qw & \qw & \qw & \ctrl{1} & \qw & \qw&\qw\\
& \gate{R_z}   &\qw    &\ccteqg
& &\qw 
& \gate{R_z(\theta_1)}  &\targ & \gate{R_z(\theta_2) }&\targ &\gate{R_z(\theta_3)} & \targ & \gate{R_z(\theta_4) } & \targ & \gate{R_z(\theta_5) } & \targ & \gate{R_z(\theta_6) }& \targ & \gate{R_z(\theta_7) } & \targ & \gate{R_z(\theta_8) } &\targ & \qw
} 
\vspace{1em}
\newcolumntype{R}{>{\raggedright\arraybackslash}p{4.5em}}%
\newcolumntype{P}{p{3.75em}}
\newcolumntype{G}{>{\columncolor{gr}}l|}
\begin{tabularx}{\linewidth}{RPPPPPPPP}
\hhline{~*{8}{-}}
\multicolumn{1}{c|}{bit 1} & \multicolumn{1}{l|}{} & \multicolumn{1}{l|}{}                         & \multicolumn{1}{l|}{}                         & \multicolumn{1}{l|}{}                         & \multicolumn{1}{G}{} & \multicolumn{1}{G}{} & \multicolumn{1}{G}{} & \multicolumn{1}{G}{} \\ \hhline{~|*{8}{-|}}
\multicolumn{1}{c|}{bit 2} & \multicolumn{1}{l|}{} & \multicolumn{1}{l|}{}                         & \multicolumn{1}{G}{} & \multicolumn{1}{G}{} & \multicolumn{1}{G}{} & \multicolumn{1}{G}{} & \multicolumn{1}{l|}{}                         & \multicolumn{1}{l|}{}                         \\ \hhline{~|*{8}{-|}}
\multicolumn{1}{c|}{bit 3} & \multicolumn{1}{l|}{} & \multicolumn{1}{G}{} & \multicolumn{1}{G}{} & \multicolumn{1}{l|}{}                         & \multicolumn{1}{l|}{}                         & \multicolumn{1}{G}{} & \multicolumn{1}{G}{} & \multicolumn{1}{l|}{}                         \\ \hhline{~*{8}{-}}
                           & $g_1$                 & $g_2$                                         & $g_3$                                                                & $g_4$                                                                & $g_5$                                                                & $g_6$                                                                & $g_7$                                         & $g_8$                                                                 
\end{tabularx}
\caption{Decomposition of a uniformly controlled $R_z$ gate with three control qubits with the three-bit Gray code that is used to find the control nodes of the CNOTs.}
\label{fig:multicontrolled3}
\end{figure}
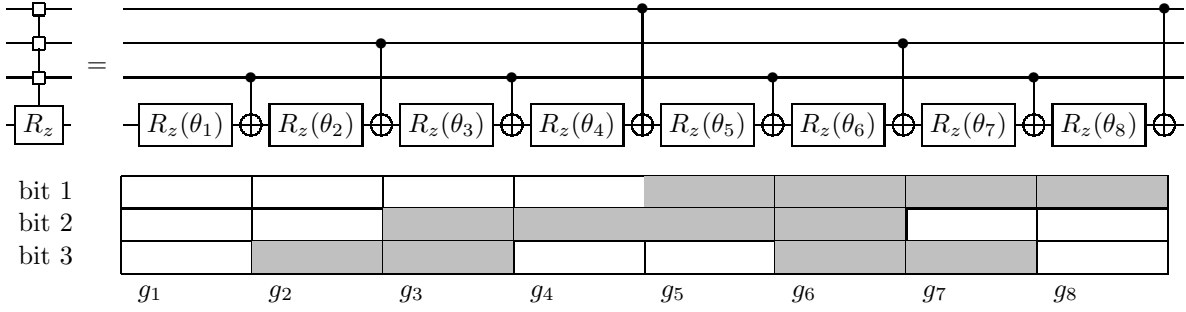

The structure of the decomposition of a uniformly controlled $R_z$ gate with three control qubits is shown in \cref{fig:multicontrolled3}.  

The $2^k$ rotation gates in the circuit each apply a rotation by some angle $\theta_j$ to the target qubit.  
The Gray code sequence means each control qubit is the control of a CNOT an even number of times, which negates some of the angles. At the same time, subsequent rotations about the same axis are additive. This means the quantum circuit is equivalent to $F_{k+1}^k(R_z(\alpha_j))$ if the angles $\theta_j$ are a solution to:

\begin{align}
    M^k \begin{bmatrix} \theta_1 \\ \vdots \\ \theta_{2^k} \end{bmatrix} &= \begin{bmatrix} \alpha_1 \\ \vdots \\ \alpha_{2^k} \end{bmatrix}
\end{align}
where the elements of matrix $M^k$ correspond to 
\begin{align}
    M_{ij}^k = (-1)^{b_{i-1}\cdot g_{j-1}}
\end{align}
where $b_i$ is the standard binary code representation of the integer $i$, and $g_j$ corresponds to the binary representation of the $j^\textrm{th}$ Gray code number. The dot denotes the bitwise inner product between the binary vectors.

\section{Full decomposition} \label{sec:decomposition}
In this section, we first introduce the basis of our decomposition: the block-ZXZ decomposition~\cite{art:blockzxz}. Then we show how to decompose the circuit into elementary gates~\cite{art:synthesisqlogic}. This decomposition method results in the same number of CNOT gates as the unoptimized quantum Shannon decomposition~\cite{art:blockzxz}. 

\subsection{Block-ZXZ decomposition} \label{subsec:blockzxz}
The proposed decomposition is based on the block-ZXZ decomposition presented in~\cite{art:blockzxz}, which shows how the method presented in~\cite{art:biunimodular} can be used to decompose a general unitary gate into the following structure:

\begin{align}
    U &= \frac{1}{2} \begin{bmatrix} A_1 & 0 \\ 0 & A_2 \end{bmatrix} \begin{bmatrix} I+B & I - B \\ I - B & I + B \end{bmatrix} \begin{bmatrix} I & 0 \\ 0 & C \end{bmatrix} \label{eq:blockzxz} \\ 
         &= \frac{1}{2} \begin{bmatrix} A_1 & 0 \\ 0 & A_2 \end{bmatrix} \left( H \otimes I \right) \begin{bmatrix} I & 0 \\ 0 & B \end{bmatrix}  \left( H \otimes I \right)  \begin{bmatrix} I & 0 \\ 0 & C \end{bmatrix} 
\end{align}
This can be represented as the following quantum circuit:

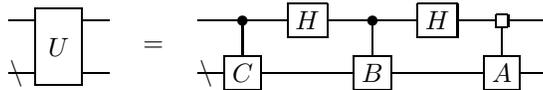
\[
\Qcircuit @C=1em @R=.7em {
 & & \multigate{1}{U} & \qw& \ccteq{1} & & \ctrl{1} & \gate{H} &\ctrl{1} & \gate{H} &\ctrlu{1} & \qw\\
& \csl & \ghost{U} & \qw& \ccteqg & \csl & \gate{C} & \qw & \gate{B} & \qw& \gate{A}  & \qw 
}
\]

To construct this circuit, we need to solve \cref{eq:blockzxz}, which requires that~\cite{art:biunimodular}: 
\begin{align}
    U &\begin{bmatrix} I \\ C^\dagger \end{bmatrix} = \begin{bmatrix} A_1 \\ A_2 \end{bmatrix} \label{eq:problem12}
\end{align}

To find matrices $A_1$, $A_2$ and $C$, we first divide the starting matrix $U$ into four equal blocks. We call the upper left block $X$, the upper right block $Y$ and the lower two blocks $U_{21}$ and $U_{22}$. This makes  $U = \begin{bmatrix} X & Y \\ U_{21} & U_{22} \end{bmatrix}$ and then use singular value decomposition to decompose $X$ and $Y$. 

For $X$, with singular value decomposition we get $X = V_X \Sigma W_X^\dagger $ with unitary matrices $V_X, W_X \in U$ and $\Sigma$ is a diagonal matrix with non-negative real numbers on the diagonal. We define $S_X = V_X \Sigma V_X^\dagger$, a positive semi-definite matrix and unitary matrix $U_X = V_X W_X^\dagger$. Then we have the polar decomposition of $X = S_X U_X$. The same method can be used to find $S_Y$ and $U_Y$ so that $Y = S_Y U_Y $. 

Then we can write
\begin{align}
    U &= \begin{bmatrix} S_X U_X & S_Y U_Y \\ U_{21} & U_{22}  \end{bmatrix}
\end{align}
and define $C^\dagger = iU_Y^\dagger U_X$ so that \cref{eq:problem12} becomes
\begin{align}
    U &\begin{bmatrix} I \\ iU_Y^\dagger U_X \end{bmatrix} = \begin{bmatrix} A_1 \\ A_2 \end{bmatrix}
\end{align}
We can find $A_1 = (S_X + iS_Y)U_X$ and $A_2 = U_{21} + U_{22}(iU_Y^\dagger U_X)$.  Finally, we rewrite \cref{eq:blockzxz} and solve for the upper left corner to get $B = 2A_1^\dagger X - I $.

\subsection{Demultiplexing} \label{sec:demultiplexing}
A gate $U = U_1 \oplus U_2$ can be decomposed into unitary matrices $V$ and $W$ and a unitary diagonal matrix $D$ so that $U = (I\otimes V) (D \oplus D^\dagger  ) (I \otimes W)$ using the method described in theorem 12 of~\cite{art:synthesisqlogic}:
\begin{align}
    \begin{bmatrix} U_1 & 0 \\ 0 & U_2 \end{bmatrix} &= \begin{bmatrix} V & 0 \\ 0 & V \end{bmatrix} \begin{bmatrix} D & 0 \\ 0 & D^\dagger \end{bmatrix} \begin{bmatrix} W & 0 \\ 0 & W \end{bmatrix} 
\end{align}
To find the values for $V$, $D$ and $W$, we first use diagonalization of $U_1 U_2^\dagger$ to get $U_1 U_2^\dagger = V D^2 V^\dagger$, where $V$ is a square matrix with columns representing the eigenvalues of $U_1 U_2^\dagger$ and $D$ a diagonal matrix whose diagonal entries are the corresponding eigenvalues. Then we can find $W$ as $W = D V^\dagger U_2$. 
The matrix $D \oplus D^\dagger$ corresponds to a multiplexed $R_z$ gate acting on the most significant qubit in the circuit. 

In a quantum circuit, demultiplexing looks like this:

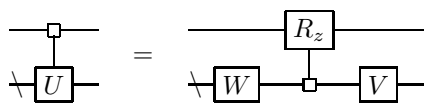
\[      
\Qcircuit @C=1em @R=.7em{
&     & \ctrlu{1}    & \qw & \ccteq{1} &      & \qw      & \gate{R_z}    & \qw       & \qw \\
&\csl    & \gate{U}  & \qw & \ccteqg   & \csl & \gate{W} &   \ctrlu{-1}  & \gate{V}  & \qw 
  } \]
      
We can use this method to demultiplex gates $A$, $B$ and $C$ from the circuit in \cref{subsec:blockzxz}, which gives the following circuit: 

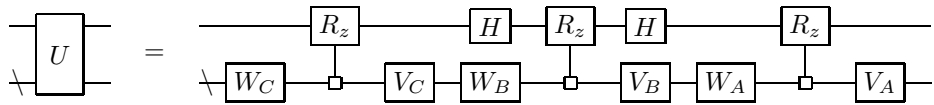
\vspace{1em}
\[
\Qcircuit @C=1em @R=.7em {
&       & \multigate{1}{U}  & \qw & \ccteq{1} & 
& \qw & \gate{R_z} & \qw & \gate{H} & \gate{R_z} & \gate{H} & \qw & \gate{R_z} & \qw & \qw \\
& \csl  &\ghost{U}          & \qw & \ccteqg   & \csl
& \gate{W_C} & \ctrlu{-1} & \gate{V_C} & \gate{W_B} & \ctrlu{-1} & \gate{V_B} & \gate{W_A}  & \ctrlu{-1} & \gate{V_A} & \qw 
}
\]
\vspace{1em}

It is clear from the circuit that gate $V_C$ can be merged with $W_B$, and that $V_B$ can be merged with $W_A$. This means we now have a circuit decomposition of an initial n-qubit gate into four (n-1)-qubit gates, three uniformly controlled $R_z$ gates and two Hadamard gates. The uniformly controlled $R_z$ gates can be decomposed as in \cref{subsec:uniformlycontrolled}. The decomposition is applied recursively to each (n-1)-qubit gate until only one-qubit gates are left, which can be decomposed using ZYZ-decomposition~\cite{art:elementarygates}. 

This leads to a total CNOT count that is the same as the unoptimized quantum Shannon decomposition~\cite{art:synthesisqlogic}: $(3/4) \cdot 4^n-(3/2) \cdot 2^n$.  

\section{Optimization} \label{sec:optimization}
Because the circuit resulting from the block-ZXZ decomposition is very similar to that of the quantum Shannon decomposition~\cite{art:synthesisqlogic}, it can be optimized using the same methods. But where QSD can merge one CNOT gate from the central $R_y$ gate, we can merge two CNOT gates into the central multiplexor. This results in a total CNOT count of $\frac{22}{48}4^n - \frac{3}{2}2^n +\frac{5}{3} = \frac{11}{24}4^n - \frac{3}{2}2^n +\frac{5}{3}$ CNOT gates for decomposing an n-qubit unitary gate. 

\subsection{Decomposition of two-qubit operators} \label{subsec:opt2q}
The decomposition can be applied recursively until the biggest blocks are the generic 2-qubit unitary gates. These can be decomposed using the optimal 3-CNOT circuit, which can be done using one of several methods~\cite{art:smaller2qcircuits,art:minimalunivtwoq,art:universalquantumcirc}. This reduces the CNOT count to $(9/16)\cdot 4^n - (3/2) \cdot 2^n$~\cite{art:synthesisqlogic}.

The CNOT count can be further reduced using the technique described in appendix A2 of~\cite{art:synthesisqlogic}. The right-most two-qubit gate can be decomposed up to the diagonal into the following circuit, which requires only two qubits~\cite{art:smaller2qcircuits}:
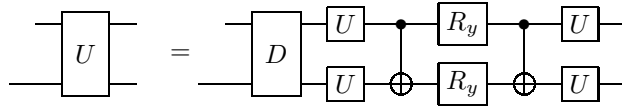
\[
\Qcircuit @C=1em @R=.7em {
 & & \multigate{1}{U} & \qw& \ccteq{1} & 
 & \multigate{1}{D} & \gate{U} &\ctrl{1} & \gate{R_y} &\ctrl{1} & \gate{U} & \qw\\
& \qw & \ghost{U} & \qw& \ccteqg & \qw 
& \ghost{\Delta} & \gate{U} & \targ & \gate{R_y} & \targ  & \gate{U} & \qw 
}
\]
The diagonal matrix can be migrated through the circuit and merged with the next two-qubit gate, which can then be decomposed and its diagonal joined with the next, until only one two-qubit gate is left. This reduces the CNOT count by $4^{n-2}-1$ gates to $(8/16)\cdot 4^n - (3/2) \cdot 2^n - 1$.

\subsection{Merging two CNOT gates into the central multiplexor} \label{subsec:optmux}
After the block-ZXZ decomposition, we first decompose only the left and right multiplexors ($A$ and $C$). We now have a circuit with two uniformly controlled $R_z$ gates. Using the decomposition of a three-qubit unitary as an example, the circuit now looks like this:
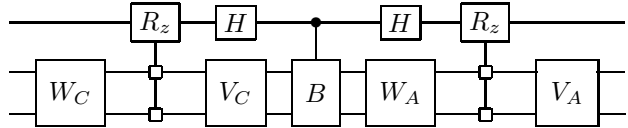
\[
\Qcircuit @C=1em @R=.7em {
& \qw & \gate{R_z} & \gate{H} & \ctrl{1} & \gate{H} & \gate{R_z} & \qw & \qw \\
& \multigate{1}{W_C} & \ctrlu{-1} & \multigate{1}{V_C} & \multigate{1}{B} & \multigate{1}{W_A}  & \ctrlu{-1} & \multigate{1}{V_A} & \qw \\
& \ghost{W_C} & \ctrlu{-1} & \ghost{V_C}  & \ghost{B} & \ghost{W_A}  & \ctrlu{-1} & \ghost{C_A} & \qw
}
\]
When decomposing the uniformly controlled $R_z$ gates, we can modify one of the decompositions so that both of the Hadamard gates are next to a CNOT:

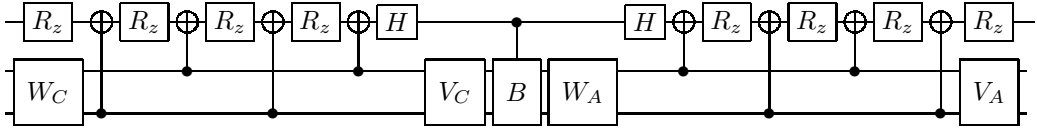
\[
\Qcircuit @C=0.3em @R=.7em {
&  \gate{R_z} & \targ & \gate{R_z} & \targ & \gate{R_z} & \targ & \gate{R_z} & \targ & \gate{H}  \qw & \qw & \ctrl{1} & \qw  & \gate{H}  & \targ 
& \gate{R_z} & \targ & \gate{R_z} & \targ & \gate{R_z} & \targ & \gate{R_z}& \qw & \qw \\
& \multigate{1}{W_C}  & \qw & \qw & \ctrl{-1} & \qw  & \qw  & \qw & \ctrl{-1} & \qw & \multigate{1}{V_C} & \multigate{1}{B} & \multigate{1}{W_A}& \qw & \ctrl{-1}   
& \qw & \qw & \qw & \ctrl{-1} & \qw & \qw & \multigate{1}{V_A} & \qw \\
& \ghost{W_C}  & \ctrl{-2} & \qw & \qw & \qw & \ctrl{-2} & \qw & \qw & \qw &  \ghost{V_C} & \ghost{B} & \ghost{W_A}  & \qw 
&  \qw & \qw & \ctrl{-2}& \qw & \qw & \qw & \ctrl{-2}  & \ghost{V_A} & \qw
}
\]

The Hadamard gates can be moved to the other side of two CNOTs, making them into CZ gates:
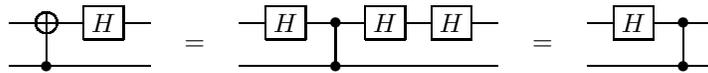
\[
\Qcircuit @C=1em @R=.7em {
& \targ     & \gate{H}  & \qw & \ccteq{1} & & \gate{H} & \control \qw & \gate{H} & \gate{H} & \qw &\ccteq{1}& & \gate{H} & \control \qw & \qw\\
& \ctrl{-1} & \qw       & \qw & \ccteqg  &  & \qw      & \ctrl{-1}    & \qw      & \qw      &  \qw &\ccteqg & & \qw & \ctrl{-1} & \qw
}
\]

This makes the circuit:
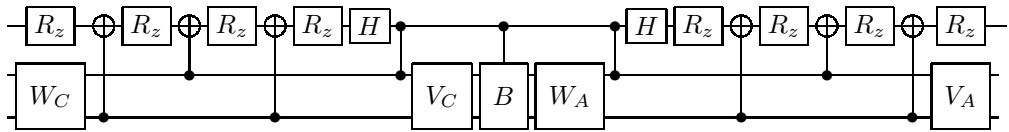
\[
\Qcircuit @C=0.3em @R=.7em {
&  \gate{R_z} & \targ & \gate{R_z} & \targ & \gate{R_z} & \targ & \gate{R_z} & \gate{H} & \control \qw & \qw & \ctrl{1} & \qw  & \control \qw & \gate{H}
& \gate{R_z} & \targ & \gate{R_z} & \targ & \gate{R_z} & \targ & \gate{R_z}& \qw & \qw \\
& \multigate{1}{W_C}  & \qw & \qw & \ctrl{-1} & \qw  & \qw & \qw & \qw & \ctrl{-1} & \multigate{1}{V_C} & \multigate{1}{B} & \multigate{1}{W_A} & \ctrl{-1}  & \qw 
& \qw & \qw & \qw & \ctrl{-1} & \qw & \qw & \multigate{1}{V_A} & \qw \\
& \ghost{W_C}  & \ctrl{-2} & \qw & \qw & \qw & \ctrl{-2} & \qw & \qw & \qw &  \ghost{V_C} & \ghost{B} & \ghost{W_A}  & \qw & \qw &  \qw & \ctrl{-2}
 &  \qw & \qw & \qw & \ctrl{-2}  & \ghost{V_A} & \qw
}
\]
The two CZ gates can be merged into the middle controlled gate $B$ together with the two (n-1)-qubit gates $V_C$ and $W_A$, similar to the optimization in appendix A1 of~\cite{art:synthesisqlogic}.

The new central gate  $\tilde B$ can be calculated as:
\begin{align}
    \tilde B = \begin{bmatrix} W_A V_C & 0 \\ 0 & (Z\otimes I)W_A\, B\,V_C (Z\otimes I) \end{bmatrix} &= %
    (CZ \otimes I) %
    \begin{bmatrix} W_A & 0 \\ 0 & W_A \end{bmatrix} %
    \begin{bmatrix} I & 0 \\  0 & B\end{bmatrix} %
    \begin{bmatrix}
        V_C & 0 \\ 
        0 & V_C \end{bmatrix} 
    (CZ \otimes I) %
\end{align}

and decomposed as a regular multiplexor, using the methods described in~\cref{sec:demultiplexing}.

\[
\Qcircuit @C=0.3em @R=.7em {
&  \gate{R_z} & \targ & \gate{R_z} & \targ & \gate{R_z} & \targ & \gate{R_z} & \gate{H} & \ctrl{1}  & \gate{H}
& \gate{R_z} & \targ & \gate{R_z} & \targ & \gate{R_z} & \targ & \gate{R_z}& \qw & \qw \\
& \multigate{1}{W_C}  & \qw & \qw & \ctrl{-1} & \qw  & \qw & \qw & \qw  & \multigate{1}{\tilde B}  & \qw 
& \qw & \qw & \qw & \ctrl{-1} & \qw & \qw & \multigate{1}{V_A} & \qw \\
& \ghost{W_C}  & \ctrl{-2} & \qw & \qw & \qw & \ctrl{-2} & \qw & \qw & \ghost{\tilde B} & \qw &  \qw & \ctrl{-2}
 &  \qw & \qw & \qw & \ctrl{-2}  & \ghost{V_A} & \qw
}
\]

This saves two CNOTs for every step of the recursion, for a total savings of $2\cdot(4^{n-2}-1)/3$ CNOT gates when stopping the recursion at generic two-qubit gates. 

\subsection{Gate count}
For a generic three-qubit unitary, the decomposition results in four generic two-qubit gates. Three of these require two CNOTs to implement, while the last one requires three CNOTs. The left and right controlled $R_z$ gates both need three CNOTs and the middle uniformly controlled $R_z$ gate requires four CNOTs to decompose. 
That makes the total CNOT count for the decomposition of a three-qubit unitary: $3\cdot 2 + 3 + 2 \cdot 3 + 4 = 19$ CNOT gates. 

To find the number of CNOT gates required for implementing bigger operators, we start with the recursive relation below. An $n$-qubit unitary requires $c_n$ CNOTs, which are at most:
\begin{align*}
    c_n &\leq 4 \cdot c_{n-1} - 3 + 3 \cdot 2^{n-1} -2  =  4 \cdot c_{n-1} + 3 \cdot 2^{n-1} -5
\end{align*}

This breaks down as follows: at each level of the recursion, the CNOT count is the sum of the CNOTs required for the decompositions of the four smaller unitaries ($c_{n-1}$) and the CNOTs required by the three quantum multiplexors ($2^{n-1}$). Three of the smaller unitaries can be implemented using one CNOT less by applying the optimization presented in \cref{subsec:opt2q}, and two of the multiplexors can be implemented using one less CNOT using the method in \cref{subsec:optmux}. 

A two-qubit unitary operator can be decomposed using at most three CNOTs ($c_2 \leq 3$), the recursive relations for 3, 4 and 5 qubit unitary operators are given below.
\begin{align*}
    c_3 &\leq 4 \cdot c_2 + 3 \cdot 2^{3-1} - 5\\
    c_4 &\leq 4 \cdot c_3 + 3 \cdot 2^{4-1} - 5 \\
    &\leq 4 \cdot 4 \cdot c_2 + 4 \cdot 3 \cdot 2^{4-2} - 4 \cdot 5  + 3 \cdot 2^{4-1} - 5\\ 
    &\leq 4^2 \cdot c_2 + 3 \cdot 2^{4-1} (4\cdot 2^{-1} + 1) - 5 \cdot (4 + 1)\\
    c_5 &\leq 4^3 \cdot c_2 + 3 \cdot 2^{5-1} (4^2 \cdot 2^{-2} + 4\cdot 2^{-1} + 1) - 5 \cdot (4^2 + 4 + 1) \\
    &\leq 4^3 \cdot c_2 + 3 \cdot 2^{5-1} (2^{2}  + 2^{1} + 1) - 5 \cdot (4^2 + 4 + 1)
\end{align*}
We can recognize the following structure~\cite{book:discretemath}:
\begin{align*}
    1 + x + x^2 + \dotsb + x^n = \frac{x^{n+1}-1}{x-1}
\end{align*}
We can use this to derive the following relation for the CNOT count for the decomposition of an $n$-qubit unitary gate:
\begin{align*}
    c_n &\leq 4^{n-2} \cdot c_2 + 3\cdot 2^{n-1} \left(\frac{2^{(n-3)+1} -1}{2-1}\right)  -  5 \left(\frac{4^{(n-3)+1} -1}{4-1}\right) \\
    c_n &\leq 4^{n-2}  \cdot c_2 + 3 \cdot 2^{n-1} (2^{n-2} -1) - \frac{5}{3} (4^{n-2} -1)\\
    c_n &\leq  \left(4^{-2}\cdot c_2 + 3 \cdot 2^{-3} - \frac{5}{3} \cdot4^{-2} \right) \cdot 4^{n}  - 3 \cdot 2^{-1} \cdot 2^{n} + \frac{5}{3}
\end{align*}
With $c_2 \leq 3$, we get the following CNOT count for the decomposition of an $n$-qubit unitary gate:
\begin{align*}
    c_n &\leq \left(\frac{3}{16}+\frac{3}{8} - \frac{5}{48}\right)\cdot 4^n - \frac{3}{2}\cdot 2^n + \frac{5}{3} \\
    c_n &\leq \frac{22}{48}\cdot 4^n - \frac{3}{2}\cdot 2^n + \frac{5}{3} 
\end{align*}

\section{Conclusion} \label{sec:conclusion}
In this paper, we presented a novel quantum decomposition method that is able to produce circuits with a gate count that is lower than existing state-of-the-art quantum decomposition methods. We used the optimizations presented by~\cite{art:synthesisqlogic}, gate commutation and gate merging to optimize the block-ZXZ decomposition~\cite{art:blockzxz}. The decomposition follows the same structure of the well-known quantum Shannon decomposition, and has the same benefit of using recursion on generic quantum gates. This means that the decomposition can take advantage of the known optimal decompositions for two-qubit unitary gates, and other small-circuit optimizations, heuristic methods or optimal decompositions for three or more qubit gates when these become available. 

Other than general unitary gates, the decomposition uses only single qubit gates and diagonal gates. This simplifies the structure and presents further opportunity for optimizations, such as accounting for specific hardware constraints like connectivity.

The circuit output of the decomposition can be compiled to any universal gateset. The resulting circuit will have an equal number of two-qubit gates when the gateset includes a two-qubit gate that is equivalent to the CNOT gate up to single qubit gates. This is the case for, among others, the  CZ gate (part of the native gateset of the IBM Heron)~\cite{art:elementarygates}, the ECR gate (IBM Eagle)~\cite{art:demonstratingQV}  and the XX gate (trapped ions)~\cite{art:trappedions}. If these circuits are executed on a quantum execution platform which has a more permissive gateset, the diagonal gates can also be implemented with uniformly controlled Z-gates~\cite{art:quantumgatedecomposer} instead of CNOTs.

As can be seen in \cref{tab:comparedecomps}, our approach improves upon the previous record holder by $(4^{n-2} -1)/3$ CNOT gates to achieve the best-known CNOT count for any generic quantum gate of size three or more qubits.
\printbibliography

\end{document}